\documentclass[conference]{IEEEtran}
\IEEEoverridecommandlockouts

\usepackage{cite}
\usepackage{amsmath,amssymb,amsfonts}
\usepackage{algorithmic}
\usepackage{graphicx}
\usepackage{textcomp}
\usepackage{xcolor}
\usepackage{url}
\usepackage{hyperref}
\usepackage{booktabs}
\usepackage{subcaption}
\def\BibTeX{{\rm B\kern-.05em{\sc i\kern-.025em b}\kern-.08em
    T\kern-.1667em\lower.7ex\hbox{E}\kern-.125emX}}

\begin{document}

\title{Physics-Aware Fluid Field Generation from User Sketches Using Helmholtz-Hodge Decomposition
}

\author{\IEEEauthorblockN{Ryuichi Miyauchi\IEEEauthorrefmark{1},
Hengyuan Chang\IEEEauthorrefmark{1}, 
\\Tsukasa Fukusato\IEEEauthorrefmark{2},  
Kazunori Miyata\IEEEauthorrefmark{1}, 
Haoran Xie\IEEEauthorrefmark{1}\IEEEauthorrefmark{3}}

\IEEEauthorblockA{\IEEEauthorrefmark{1}Japan Advanced Institute of Science and Technology, 
\IEEEauthorrefmark{2}Waseda University}}

\maketitle
\newcommand\blfootnote[1]{%
  \begingroup
  \renewcommand\thefootnote{}\footnote{#1}%
  \addtocounter{footnote}{-1}%
  \endgroup
}
\blfootnote{\IEEEauthorrefmark{3}Corresponding author (xie@jaist.ac.jp).} 

\begin{abstract}
Fluid simulation techniques are widely used in various fields such as film production, but controlling complex fluid behaviors remains challenging. While recent generative models enable intuitive generation of vector fields from user sketches, they struggle to maintain physical properties such as incompressibility. To address these issues,
this paper proposes a method for interactively designing 2D vector fields. Conventional generative models can intuitively generate vector fields from user sketches, but remain difficult to consider physical properties. Therefore, we add a simple editing process after generating the vector field.

In the first stage, we use a latent diffusion model~(LDM) to automatically generate initial 2D vector fields from user sketches.
In the second stage, we apply the Helmholtz-Hodge decomposition to locally extract physical properties such as incompressibility from the results generated by LDM and recompose them according to user intentions. Through multiple experiments, we demonstrate the effectiveness of our proposed method.
\end{abstract}


\section{Introduction}
Fluid simulation techniques are extensively utilized in film production; however, the complexity of fluid dynamics poses a barrier for users manipulating the flow movements.
Against this background, many researchers have discussed how to intuitively control fluid flows~\cite{wang_physics-based_2024}. 
One solution is to automatically estimate vector fields from user-drawn sketches by using generative models~\cite{hu_sketch2vf_2019, xie2024dualsmoke, chang_diffsmoke_2025}.
This approach does not require complex parameter tuning, but these methods are unsuitable for considering physical properties such as incompressibility. 

\begin{figure*}[h]
    \centering
    \includegraphics[width=0.95\linewidth]{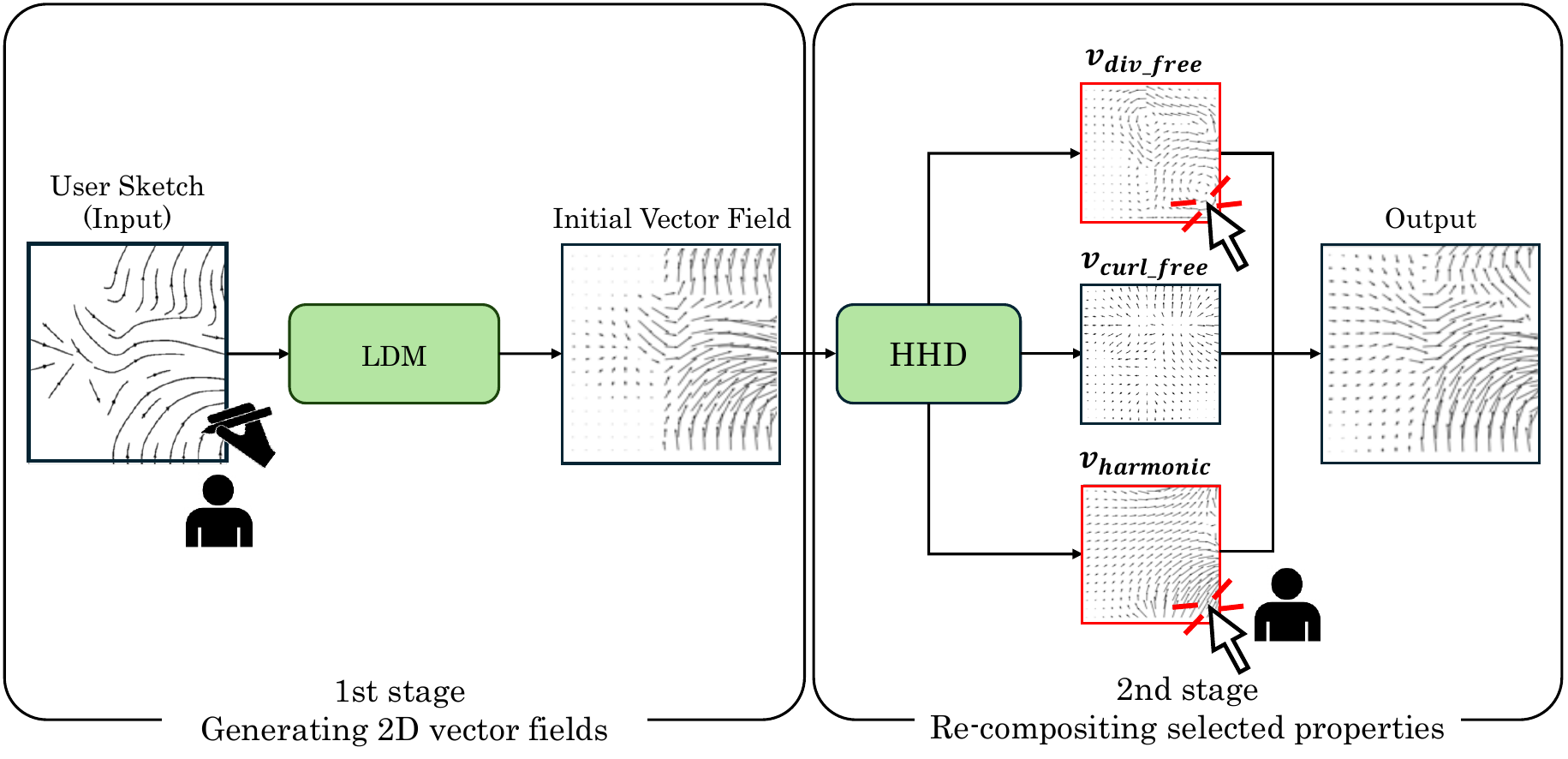}
    \caption{Our framework. \textcolor{black}{Our system initially generates a vector field from user sketch input using LDM. Subsequently, the Helmholtz-Hodge decomposition-based vector field editing interface enables users to extract desired physical properties from the initial vector field (which will be described in more detail in Section~\ref{sec:ui}).  In this example, users obtain a physically plausible vector field by decomposing the  vector field into the divergence-free component ($\mathbf{v}_{\text{div-free}}$) and harmonic component ($\mathbf{v}_{\text{harmonic}}$) with Helmholtz-Hodge decomposition.}}
    \label{fig:overview}
\end{figure*}

To address these issues, we introduce a two-stage approach to interactively design/edit 2D vector field with physical properties. The framework of the proposed method is illustrated in \autoref{fig:overview}. In the first-stage, by referring to existing methods, we construct a latent diffusion model~(LDM) to generate 2D vector field from a user-drawn sketch. 
In the second-stage, we apply the Helmholtz-Hodge decomposition into the generated fields to locally extract three properties: (i)~irrotational vector fields of potential flow, (ii)~incompressible vector fields of divergence-free flow, and (iii)~harmonic components, from user-specified regions, and re-composite the user-selected properties interactively. 
This method enables users to design a vector field while considering physical properties intuitively and efficiently. 

The main contributions of our work are as follows:

\begin{itemize}
\item We propose a novel two-stage approach that combines a latent diffusion model and the Helmholtz-Hodge decomposition for designing 2D vector fields.

\item We develop an interactive interface that allows users to locally extract and control physical properties while maintaining their intended flow designs.

\item We demonstrate that our method can generate physically plausible vector fields without complex parameter tuning, through various experiments and applications.
\end{itemize}

\section{Related Works}

\subsection{Keyframe-based Fluid Control}
Gregson et al.~\cite{gregson_capture_2014} demonstrated that by using the pressure solver as a proximity operator, vector fields that simultaneously satisfy consistency and incompressibility conditions from fluid sequence data (e.g., smoke density field sequences) can be computed on a proximity optimization framework.

Pan et al.~\cite{pan_efficient_2017} proposed a method to compute a dense sequence of control force fields that can drive the smoke shape to match several keyframes at certain time instances.
Although this approach enables users to directly control the smoke shape instead of careful parameter tuning, it requires a large number of keyframes to control the fluid movements.

\subsection{Sketch-based Fluid Control}
EnergyBrushes~\cite{xing_energy-brushes_2016} is a system that allows users to specify the movement of flow elements with physical properties through sketch operations. 
By using this system, users can intuitively design fluid flows with artistic creativity while maintaining physical consistency.
However, users must carefully check the location and speed of each flow element when designing large-scale and complex flows.

To address these challenges, Hu et al.\cite{hu_sketch2vf_2019} proposed a cGAN-based method to generate 2D vector fields from a user-drawn sketch. In addition, Chang et al.~\cite{chang_diffsmoke_2025} extended Hu's method to improve the qualities of vector fields by using LDM model. These results demonstrate that generative models serve as effective tools for designing vector fields from user sketches. However, such approaches still struggle to account for physical properties such as incompressibility,  even if the user carefully sketches the fluid flows. Therefore, we aim to consider physical properties of vector fields in contrast to these previous approaches.


\section{Proposed Method}
 Our approach introduces a two-stage method: (1) generation of initial 2D vector fields using a sketch-based LDM for global flow design, and (2) a user interface for vector field editing that applies the Helmholtz-Hodge decomposition to extract desired physical properties. In this section, we first describe our dataset generation process, then detail the network architecture, and finally present the methodology and user interface for vector field editing.

\subsection{Dataset Generation}
\label{sec:dataset}

\begin{figure}[htbp]
    \centering
    \begin{subfigure}[b]{1.0\linewidth}
        \centering
        \includegraphics[width=\linewidth]{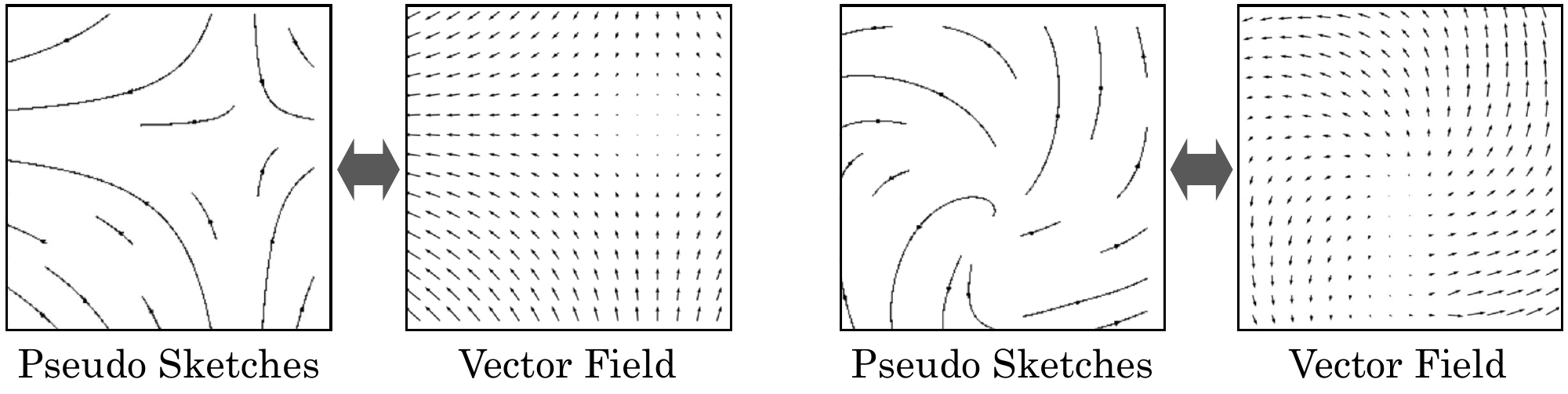}
        \caption{linear combination of basic patterns.}
        \label{fig:subfig-a}
    \end{subfigure}

    \vspace{10pt} 

    \begin{subfigure}[b]{1.0\linewidth}
        \centering
        \includegraphics[width=\linewidth]{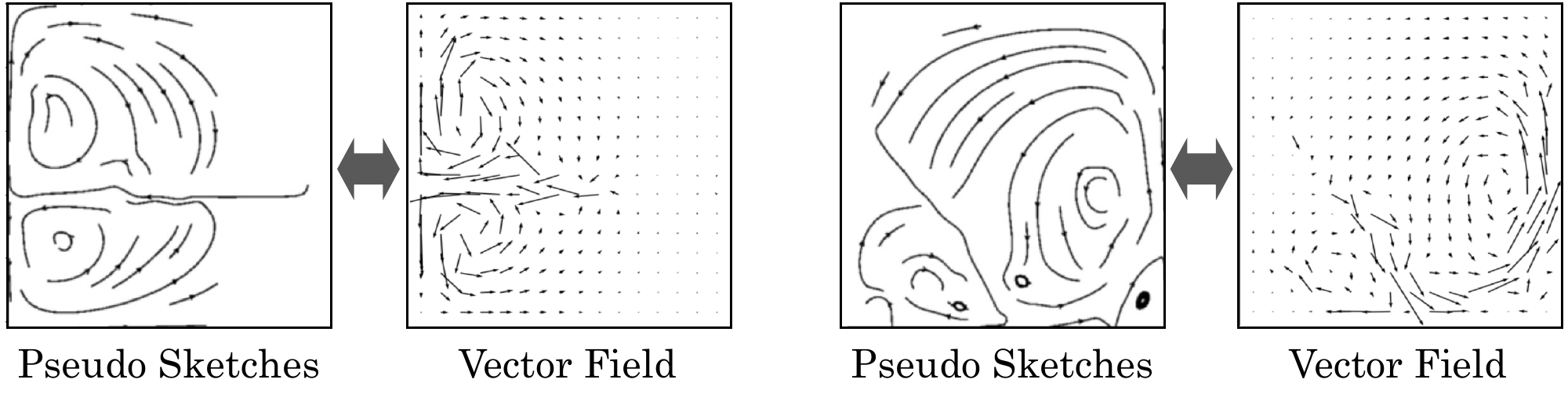}

        \caption{fluid simulator.}
        \label{fig:subfig-b}
    \end{subfigure}

    \vspace{10pt} 

    \begin{subfigure}[b]{1.0\linewidth}
        \centering
        \includegraphics[width=\linewidth]{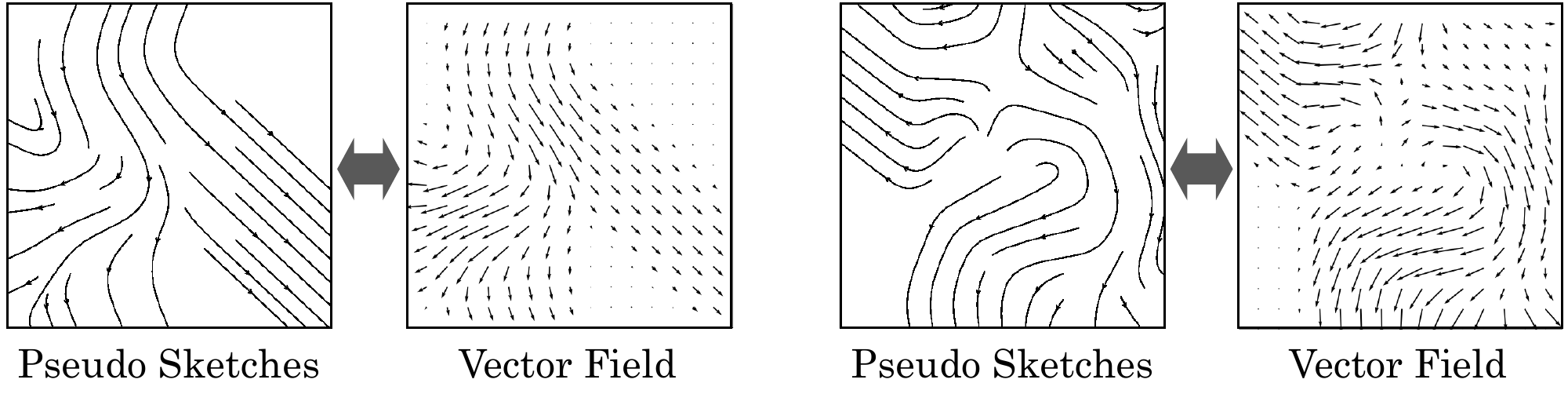}
        \caption{rule-based method.}
        \label{fig:subfig-b}
    \end{subfigure}
    
    \caption{Examples of pairwise data of pseudo sketches and 2D vector fields. We provide data examples generated with the three strategies mentioned above. For each pair of data, the left is extracted pseudo sketches, the right is the vector field.}
    \label{fig:dataset-samples}
\end{figure}

In our problem setting, training the network requires pairs of user sketches and corresponding fluid flows. However, such pairwise data is challenging to obtain in the real world. Therefore, we used synthesized sketch-like data, named pseudo-sketches, from 2D vector fields.

In order to generate 2D vector fields, we employed three approaches. 
One approach is to linearly combine some basic patterns of 2D vector fields. In our implementation, we prepared convergent, divergent, vortex, constant vector, randomly rotated constant vector, saddle point, and sine waves patterns, and synthesized a total of 16,558 vector fields, including simple and complex fields.\autoref{fig:dataset-samples}(a) shows examples of the synthesized results.

The second approach is to use existing fluid simulators based on the incompressible Navier-Stokes equations. 
First, we pre-generated a variety of inflow scenarios in which fluid flows out from circular regions, by adjusting the radius, position, and flow direction of each region. 
Next, we performed 150 simulations under each scenario and synthesized a total of 5,639 vector fields. 
%
Note that we used PhiFlow library\footnote{\url{https://tum-pbs.github.io/PhiFlow/}}  
and the semi-Lagrangian method for advection calculations. 
\autoref{fig:dataset-samples}(b) shows examples of the generated results.

The third approach is based on a simple rule.  
We select one direction of main flows from four candidates (i.e., up, down, left, or right) and generate 2D vector fields while changing the meandering degree, width, and intensity of main flows. 
To obtain various vector fields both globally and locally, we randomly set (1)~parameters of tributaries that branch from the main fluid flow (i.e., locations, direction, and spread), (2)~vortex's location and intensities, and (3)~source~/~sink locations, scales, and intensities.
In addition, to improve the smoothness, we applied Gaussian filtering and normalized each vector. 
This method can simply handle the local variations compared with the first approach, and the irrotational flow components compared with the second approach. 
We synthesized a total of 3,000 vector fields for training and 1,200 for evaluation. \autoref{fig:dataset-samples}(c) shows examples of the generated results. 
These are equally divided into six patterns, with 500 samples per pattern in the training data and 200 samples per pattern in the evaluation data. 
The patterns are (i)~vector fields containing only irrotational components, (ii)~vector fields containing only incompressible components, (iii)~vector fields containing only harmonic components, (iv)~vector fields containing both irrotational and harmonic components, (v)~vector fields containing both incompressible and harmonic components, and (vi)~vector fields containing all components. 
This dataset comprehensively covers both global and local physical properties, making it suitable for training and evaluation purposes.

The generated vector fields are 2-channel data normalized by norm and Z-score scaling with the resolutions $256\times256$.
From the generated 2D vector field, we synthesized the corresponding pseudo sketches, which are binarized 1-channel data with the resolutions $256\times256$, based on the StreamPlot function in matplotlib.

\subsection{Sketch-Conditioned Vector Field Generation}

\label{sec:ldm}
\begin{figure}[]
    \centering
    \includegraphics[width=1\linewidth]{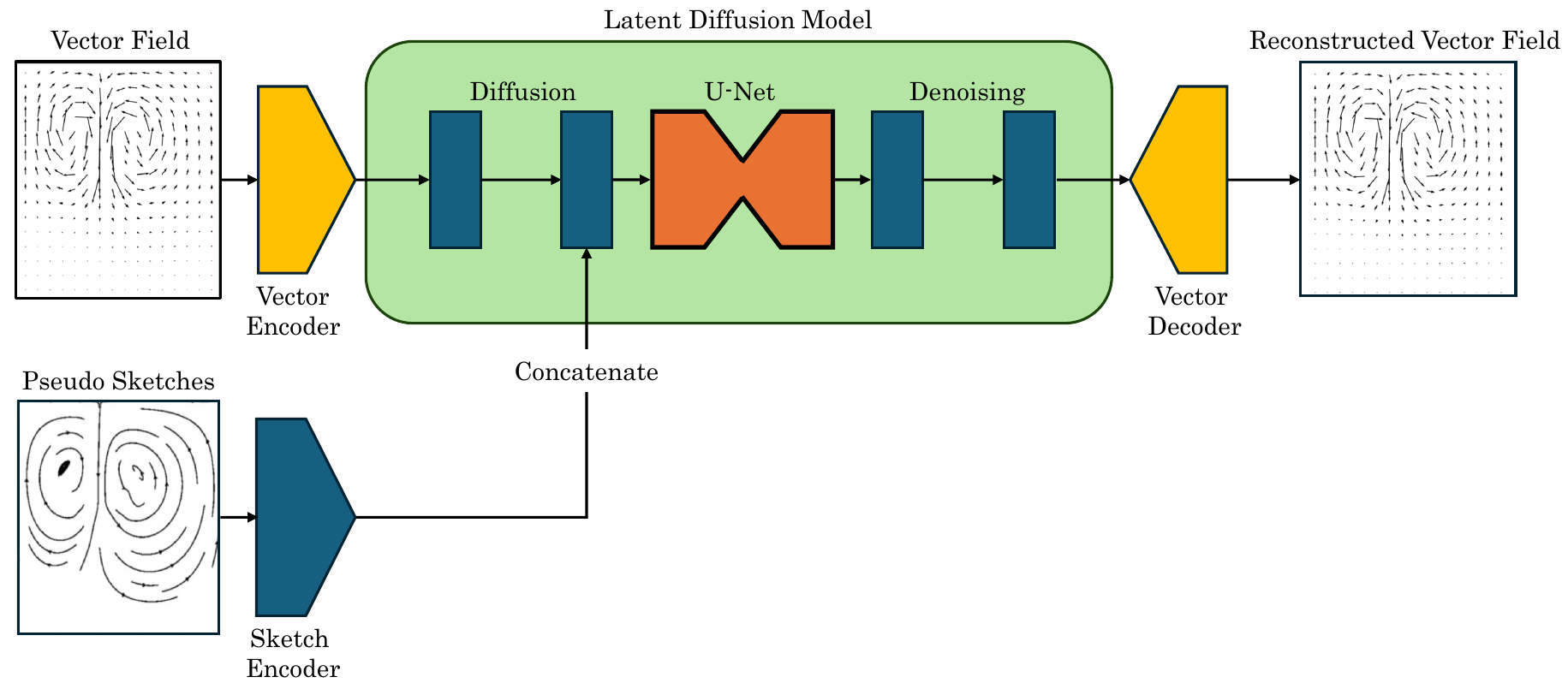}
    \caption{Architecture diagram of the latent diffusion model for generating 2D vector fields from sketches. \textcolor{black}{The encoded latent features of vector fields are transformed into standard Gaussian noise with zero mean and unit variance through the diffusion process. This Gaussian noise is concatenated with encoded sketch latent features for denoising processing. In the denoising process, U-Net~\cite{ronneberger_u-net_2015} predicts the noise added at each timestep. During inference, the final vector field is reconstructed by sequentially removing the predicted noise at each timestep with the guidance of encoded sketch latent features.}}
    \label{fig:ldm}
\end{figure}
The latent diffusion model~\cite{rombach_high-resolution_2022} is a high-quality image generation method that reduces computational costs by using auto-encoders. The auto-encoder removes perceptually meaningless high-frequency information in input data and compresses data into a latent space.
We apply the latent diffusion model to develop a method for generating 2D vector fields from sketches. Our system consists of an encoder for feature extraction from sketches, an encoder for feature extraction from vector fields, a diffusion model that transforms between these two feature representations, and a vector decoder that converts the generated feature representations into 2D vector fields (see \autoref{fig:ldm}). During the transformation process, feature representations are learned through a latent space of 2 channels and $64\times64$ dimensions.

The sketch auto-encoder consists of an encoder and a decoder. The encoder converts input sketches into 2D latent representations. It processes through convolutional layers, batch normalization layers, and SE-ResNet blocks in sequence, gradually transforming a $256\times256$ input image into a $2\times64\times64$ feature map. During this process, the number of feature channels gradually increases from 1 ($16\rightarrow32\rightarrow64\rightarrow128\rightarrow256\rightarrow512$) and finally becomes 2 channels. In each SE-ResNet block~\cite{hu_squeeze-and-excitation_2018}, the Squeeze-and-Excitation mechanism adaptively adjusts the importance of features. The decoder has a symmetric structure to the encoder and reconstructs the original sketch from the latent representation through transposed convolutional layers, batch normalization layers, and SE-ResNet blocks. The final layer normalizes the output using a Sigmoid function. The mean squared error (MSE) is used for evaluating training loss.

The vector field auto-encoder has a similar structure to the sketch auto-encoder. The only difference exists in the final layer, where the output is the result of the transposed convolution. MSE was used as the loss function.
For the latent diffusion model, we employed the sketch-conditioned generative model proposed by Chang et al.~\cite{chang_diffsmoke_2025}.


\subsection{Vector Field Editing}
The output vector fields generated by the latent diffusion model continue to struggle with incorporating the physical properties desired by users. 
To solve the problem, we apply the Helmholtz-Hodge decomposition to the LDM-based vector fields, thereby editing them to account for physical properties such as incompressibility and irrotationality.

\vspace{2mm}
\subsubsection{Helmholtz-Hodge Decomposition}

The Helmholtz-Hodge decomposition is a method for decomposing an input 2D vector field~$\mathbf{v}$ into its divergence-free component~$\mathbf{v}_{\text{div-free}}$, curl-free component~$\mathbf{v}_{\text{curl-free}}$, and harmonic component~$\mathbf{v}_{\text{harmonic}}$ through the following steps. 
For the divergence-free component~$\mathbf{v}_{\text{div-free}}$, we solve the Poisson equation for the stream function~\(\psi\) using the vorticity~\(\alpha\) of the input vector field:

\begin{align}
\alpha &= \frac{\partial v}{\partial x} - \frac{\partial u}{\partial y}\\[1em]
L\psi &= -\alpha\\[1em]
\mathbf{v}_{\text{div-free}} &= \left(\frac{\partial \psi}{\partial y},-\frac{\partial \psi}{\partial x}\right)
\label{eq:div-free}
\end{align}
\noindent
where $(u,v)$ are the $x$ and $y$ components of the input 2D vector field, and $L$ is the 2D Laplacian operator. 
Note that in this paper, the Poisson equation was solved using the conjugate gradient method.

For the curl-free component~$\mathbf{v}_{\text{curl-free}}$, we calculate it by solving the Poisson equation for the scalar potential~\(\phi\) using the divergence component~\(\beta\) of the input field:

\begin{align}
\beta &= \frac{\partial u}{\partial x}+\frac{\partial v}{\partial y}\\[1em]
L\phi &= \beta \\[1em]
\mathbf{v}_{\text{curl-free}} &= \nabla \phi
\end{align}
%

The harmonic component~$\mathbf{v}_{\text{harmonic}}$ is calculated by subtracting the divergence-free and curl-free components obtained above from the original vector field:

\begin{align}
\mathbf{v}_\text{harmonic} &= \mathbf{v}-(\mathbf{v}_\text{div-free}+\mathbf{v}_\text{curl-free})   
\end{align}

The Helmholtz–Hodge decomposition was implemented in Python using the SciPy library and executed on an i7-13700KF CPU.
For $256\times256$ and $64\times64$ vector fields, the average processing times were approximately 0.45 seconds and 0.02 seconds, respectively, based on 10 runs for each resolution.

\vspace{2mm}
\subsubsection{User Interface}
\label{sec:ui}

\begin{figure}
    \centering
    \includegraphics[width=0.8\linewidth]{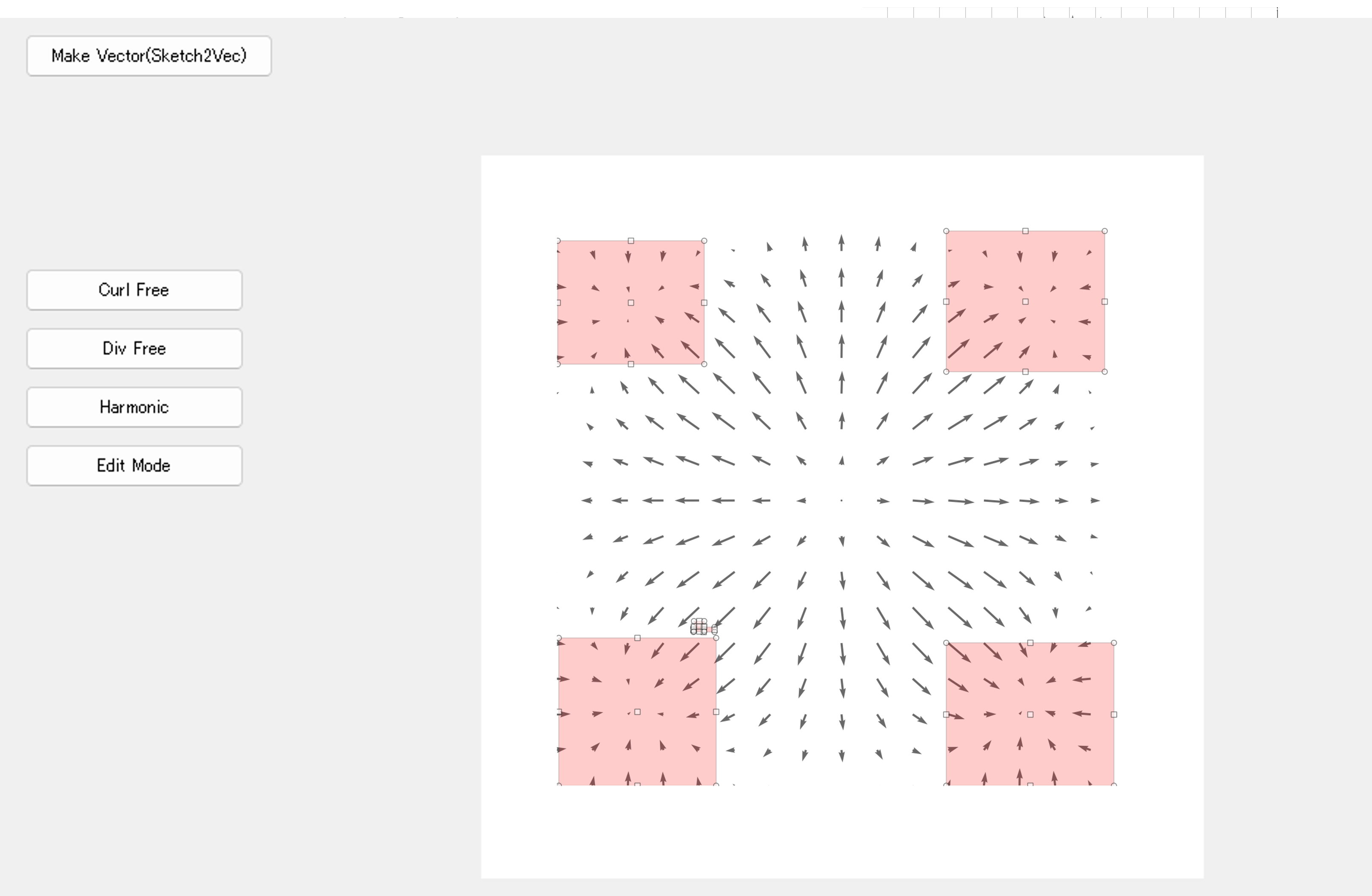}
    \caption{The screenshot of our user interface.}
    \label{fig:UI}
\end{figure}
We implemented a user interface for modifying 2D vector fields generated from user-drawn sketches, as shown in \autoref{fig:UI}.

First, users specify rectangular areas (in red) on the 2D vector field by enabling ``Edit Mode.'' For the selected regions, the system enables users to interactively extract physical properties and re-composite them through the following operations:

\vspace{2mm}
\begin{itemize}
\item \textbf{Curl Free}: Extracting the irrotational component from the selected region.

\item \textbf{Divergence Free}: Extracting the incompressible component from the selected region.

\item \textbf{Harmonic}: Extracting the harmonic component computed as the residual from the selected region.

\end{itemize}
\vspace{2mm}
Of course, users can select and recompose multiple extracted properties by clicking multiple buttons. For example, by clicking both ``Curl Free'' and ``Divergence Free'' buttons, the harmonic component only is removed from the selected region.


\section{Results}
\begin{figure}[b]
    \centering
        \includegraphics[width=0.8\linewidth]{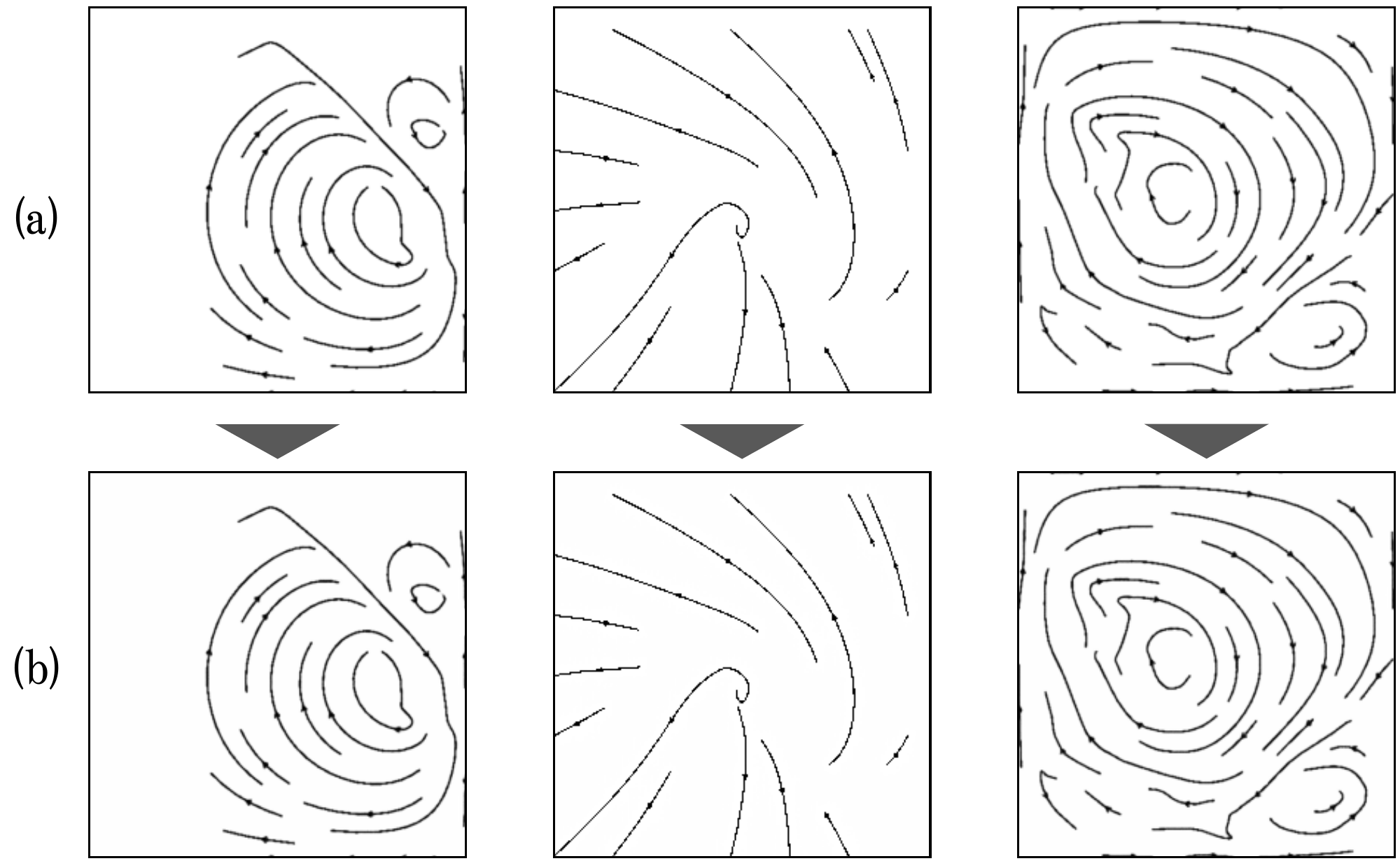}
    \caption{Examples of reconstruction results from the sketch auto-encoder. (a) pseudo sketch (input) and (b) generated results. These results show that the auto-encoder successfully preserves the line placement, curvature, and overall flow structures with high fidelity.}

    \label{fig:sketch_reconstruction}
\end{figure}

\begin{figure}[t]
    \centering
        \includegraphics[width=0.8\linewidth]{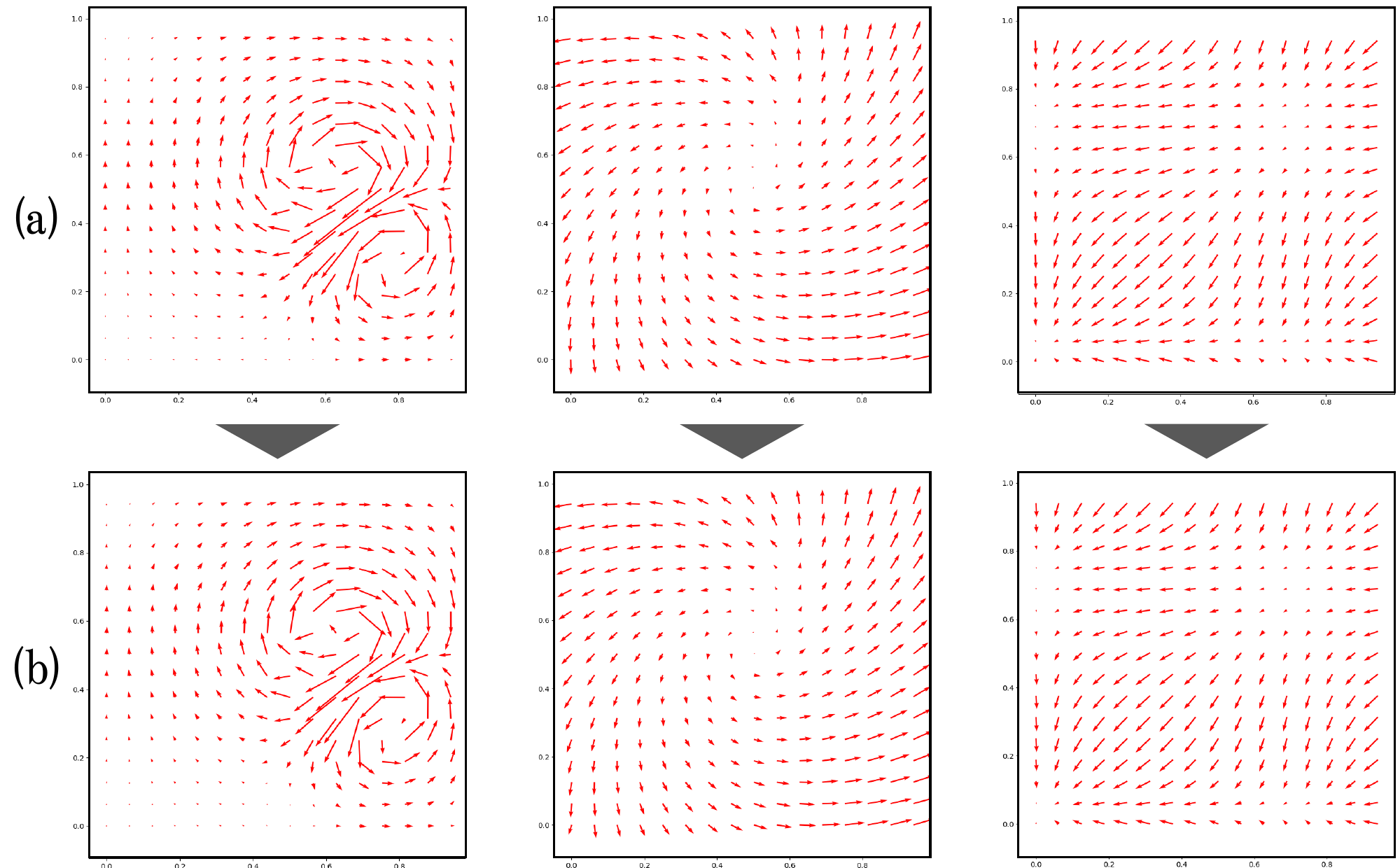} \\
    \caption{Examples of reconstruction results from the vector field auto-encoder. (a) Input vector field and (b) generated results.}
    \label{fig:vector_field_reconstruction}
\end{figure}


\subsection{Training Results}
We first synthesized the dataset with the first and second approaches in \autoref{sec:dataset} to train the sketch auto-encoder and vector field auto-encoder. The Adam optimizer was trained with 150 epochs and a batch size of 64. Subsequently, we added 3,000 new data samples that generated by the third method. We retrained the vector field auto-encoder for 76 epochs and the sketch auto-encoder for 21 epochs. The final loss values were 0.000840 for the vector field auto-encoder and 0.0000459 for the sketch auto-encoder. 

\autoref{fig:sketch_reconstruction} and \autoref{fig:vector_field_reconstruction} show the reconstruction results from the sketch auto-encoder and vector field auto-encoder, respectively. We sampled the vector field with an interval of 16 to achieve the vector field visualization.

From these results. We confirm that the sketch auto-encoder achieves high-quality reconstruction results, and the differences from the pseudo sketches~(input) are visually indistinguishable. Similarly, the vector field auto-encoder successfully reconstructs the data, since the direction and magnitude of each vector match between input and output, and the detailed features can be represented visually and numerically.

\begin{figure}
    \centering
     \includegraphics[width=0.8\linewidth]{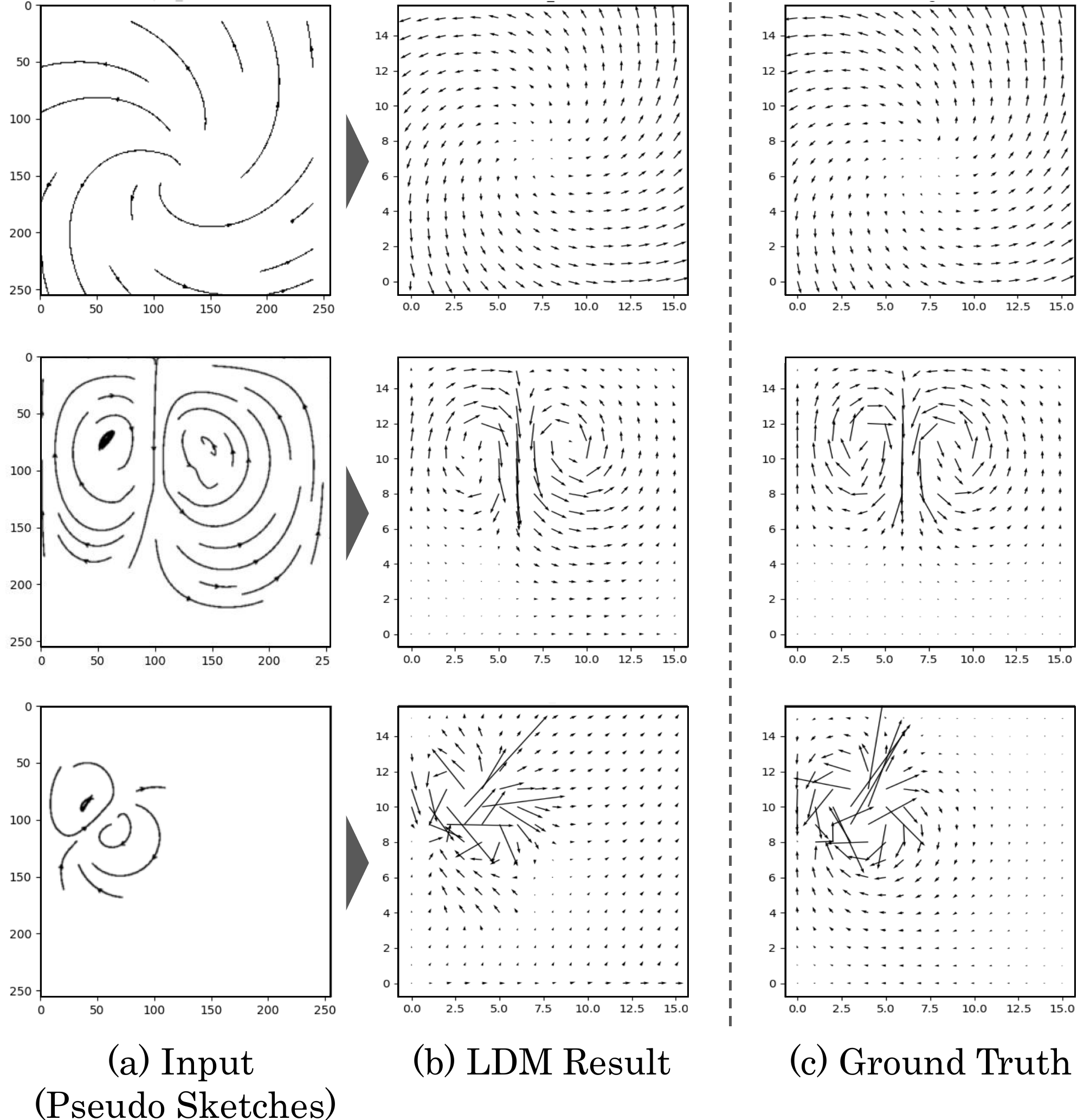}
    \caption{Comparison between 2D vector fields generated from pseudo sketches and ground truth. The shape and direction of generated flows match the ground truth.}
    \label{fig:LDM_recon}
\end{figure}

The latent diffusion model was trained on the same dataset following the training of SE-ResNet-based auto-encoder (see subsection III-B). Pre-trained weights obtained from preliminary experiments were employed as initial values to reduce the training time. The Adam optimizer was utilized for training over 85 epochs with a batch size of 32. Furthermore, we used 3,000 additional data samples generated using the third method for 74 epochs of additional training. The final loss value was 0.001070. The reconstruction results are shown in \autoref{fig:LDM_recon}.
For quantitative evaluation of results (B) and (C), we evaluated using 1,200 test data samples generated by the third method. However, the previous 74 epochs of training did not achieve sufficient generation performance for this test dataset. Therefore, we conducted additional training for 103 epochs using only the aforementioned 3,000 additional data samples from the third method.

We conducted training and inference using a computer with RTX 4090 GPU.
The training time is above five days.
For inference, we performed 32 prediction tasks and averaged the total time, resulting in approximately 3.19 seconds per sample for vector field generation.

\begin{table}[t]
\centering
\caption{Quantitative evaluation of LDM-generated vector fields for different physical properties. We evaluate reconstruction accuracy using MSE, RMSE, and SSIM across six types of vector fields and their average.}
\label{tab:ldm_reconstruction}
\begin{tabular}{l|ccc}
\toprule
Dataset & MSE~$\Downarrow$ & RMSE~$\Downarrow$ & SSIM~$\Uparrow$ \\
\midrule
All Components & 1.220 & 0.917 & 0.453 \\
Harmonic & 1.121 & 0.858 & 0.596 \\
Incompressible & 0.819 & 0.779 & 0.422 \\
Incompressible-Harmonic & 1.116 & 0.875 & 0.576 \\
Irrotational & 0.067 & 0.244 & 0.638 \\
Irrotational-Harmonic & 0.510 & 0.600 & 0.715 \\
Average & 0.809 & 0.712 & 0.567 \\

\bottomrule
\end{tabular}
\end{table}

\subsection{Evaluation of Vector Field Reconstruction Using LDM}
\begin{figure}
\centering
\includegraphics[width=0.9\linewidth]{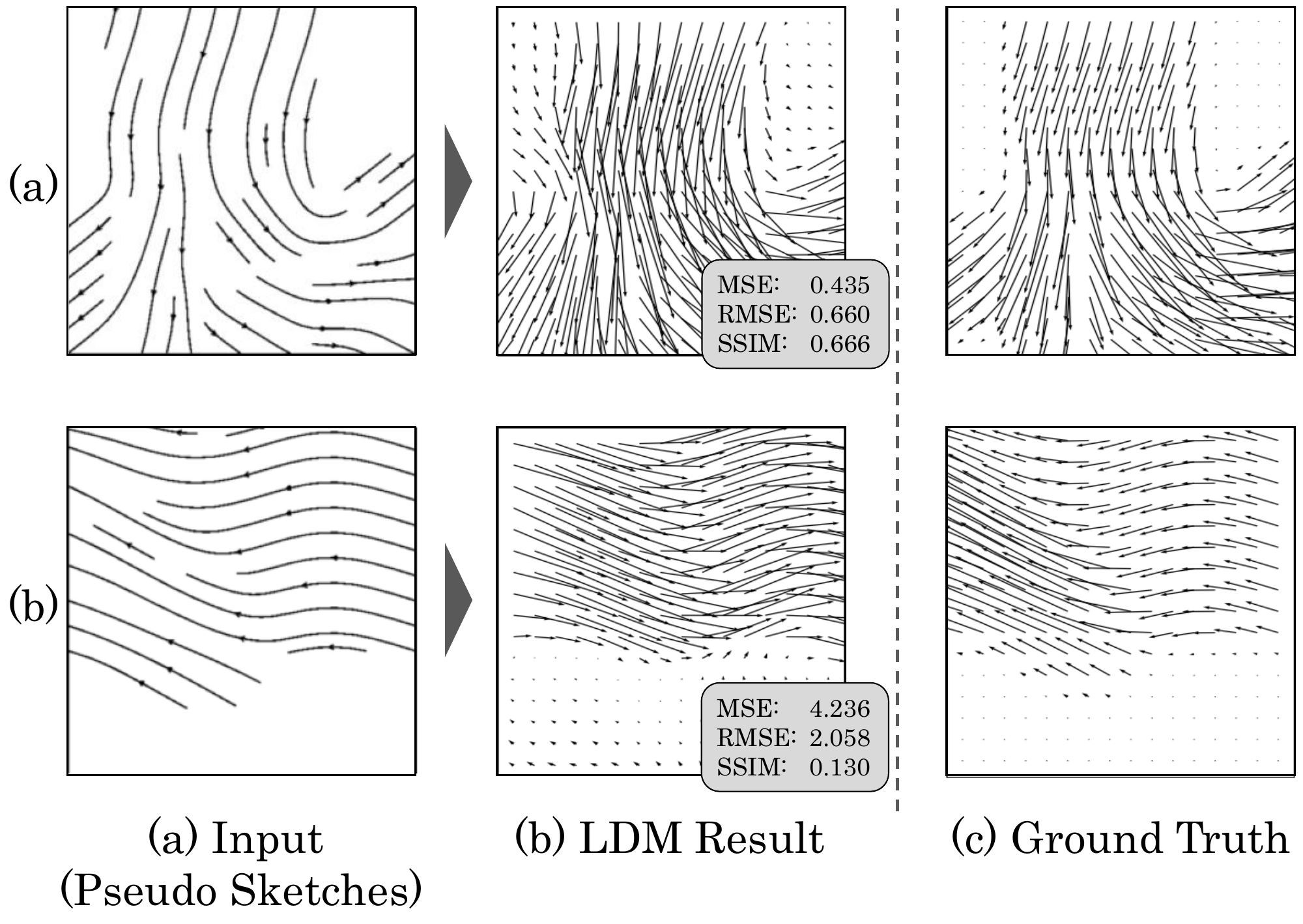}
\caption{Limitations of LDM-based vector field generation. (a) An example where controlling the vector magnitude using sketches is challenging. While the flow direction follows the sketch, differences in vector magnitudes result in a high loss. (b) A case where the generated vector field has reversed directions. This leads to significantly high loss values, with MSE reaching 4.236 and SSIM showing a low value of 0.1307.}
\label{fig:ldm_lim}
\end{figure}

We evaluated the performance of our trained LDM by quantitatively comparing the reconstructed vector fields with the ground truth vector fields.

For quantitative evaluation, we employed MSE (Mean Squared Error), RMSE (Root Mean Squared Error), and SSIM (Structural Similarity Index Measure)~\cite{wang_image_2004} as evaluation metrics. MSE and RMSE measure the error between ground truth and reconstructed data to evaluate reconstruction accuracy. SSIM generates grayscale images based on the vector field norm and measures structural similarity by weighting them with the average of cosine similarities to account for vector field directions.

\autoref{tab:ldm_reconstruction} shows that the irrotational vector field dataset achieved the best performance, with an MSE of 0.067 and an RMSE of 0.244. In contrast, other datasets exhibited higher MSE values, particularly the dataset containing all components where the MSE increased to 1.220. The average SSIM was 0.567, and the dataset containing both irrotational and harmonic components maintained a relatively high SSIM value of 0.715.
The results and vector field visualizations reveal two main limitations in LDM performance. First, as shown in \autoref{fig:ldm_lim}(a), controlling vector magnitudes through sketches proves challenging. Although the LDM-generated flow fields effectively capture the flow characteristics indicated by sketches, the MSE and RMSE values remain high due to the absence of magnitude-representing features in the pseudo-sketches. Second, as illustrated in \autoref{fig:ldm_lim}(b), the model exhibits a tendency to reconstruct vector fields with reversed directions. This limitation is particularly pronounced in vector fields containing rotational or harmonic components, resulting in increased MSE and RMSE values while reducing the SSIM.

\subsection{Evaluation of the Proposed Method}
\begin{table}[t]
\centering
\caption{Evaluation results of irrotational vector fields: LDM-based fields and editing fields with the Helmholtz-Hodge decomposition. MSE, RMSE, and SSIM evaluate overall accuracy, while VPE and CME assess physical properties specific to irrotational fields. $\downarrow$ indicates lower scores are better, $\uparrow$ for the other case.}
\label{tab:physical_evaluation_curlfree}
\begin{tabular}{l|ccccc}
\toprule
Method & MSE~$\Downarrow$ & RMSE~$\Downarrow$ & SSIM~$\Uparrow$ & VPE~$\Downarrow$ & CME~$\Downarrow$ \\
\midrule
LDM+HHD & \textbf{0.036} & \textbf{0.176} & \textbf{0.763} & \textbf{89.779} & \textbf{0.000} \\
LDM-only & 0.063 & 0.237 & 0.652 & 89.822 & 0.009 \\  
\bottomrule
\end{tabular}
\end{table}

\begin{table}[t]
\centering
\caption{Evaluation results of incompressible vector fields: LDM-based fields and editing fields with the Helmholtz-Hodge decomposition. MSE, RMSE, and SSIM evaluate overall accuracy, while SFE and CS assess physical properties specific to incompressible fields. $\downarrow$ indicates lower scores are better, $\uparrow$ for the other case.}
\label{tab:physical_evaluation_divfree}
\begin{tabular}{l|ccccc}
\toprule
Method & MSE~$\Downarrow$ & RMSE~$\Downarrow$ & SSIM~$\Uparrow$ & SFE~$\Downarrow$ & CS~$\Downarrow$ \\
\midrule
LDM+HHD & \textbf{0.753} & \textbf{0.735} & \textbf{0.451} &  \textbf{1895.659} & \textbf{0.000} \\
LDM-only & 0.794 & 0.770 & 0.432 & 1895.912 & 0.00001 \\

\bottomrule
\end{tabular}
\end{table}
To evaluate the effectiveness of our method, we compare LDM-based results with final vector fields edited by our system. This comparison assesses how well our method achieves both user-intended flow patterns and desired physical properties.
For evaluation data, we used 200 irrotational vector fields and 200 incompressible vector fields generated using the last approach described in \autoref{sec:dataset}. For each vector field, we compare the generated results with the ground truth data using MSE,RMSE, and SSIM weighted by the average of cosine similarities for quantitative and visual evaluation. 
Additionally, for physical property evaluation, we employed Curl Magnitude Error (CME, the average L2 norm of vorticity magnitude) and Vector Potential Error (VPE) for irrotational vector fields, and Continuity Score (CS)~\cite{cao_teaching_2024} and Stream Function Error (SFE)~\cite{farneback_two-frame_2003} for incompressible vector fields. This comprehensive set of metrics allows us to evaluate both the flow design capability and physical property control of our method.

\begin{figure}[t] 
\centering 
\includegraphics[width=1\linewidth]{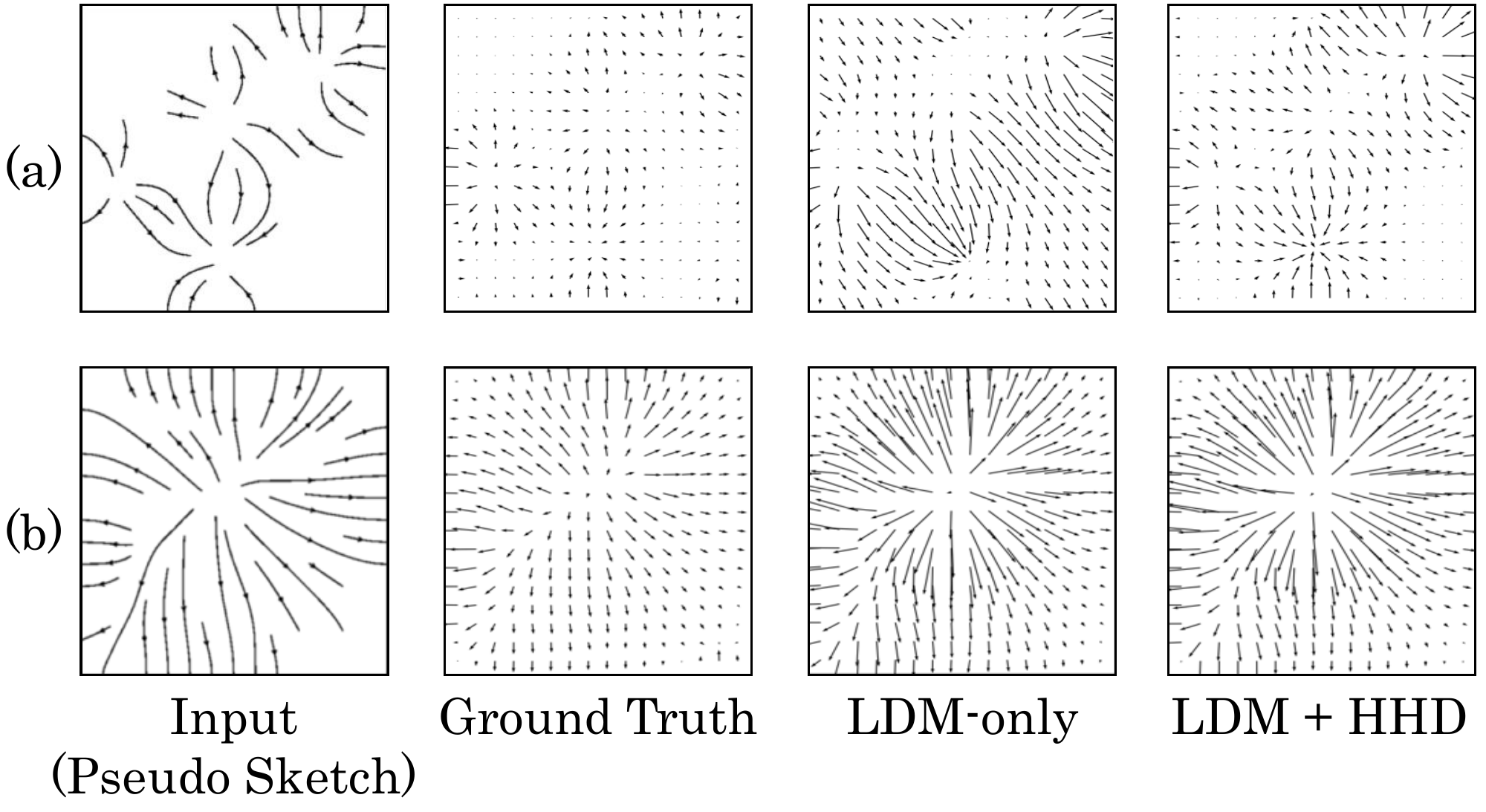} 
\caption{Examples of irrotational vector field generation and decomposition results: (a) An example where unnecessary harmonic components are removed. By extracting the curl-free part using Helmholtz-Hodge decomposition (HHD), SSIM significantly improves from 0.2371 to 0.6554. (b) An example with a significantly high VPE value. Although the flow direction of the generated vector field matches the ground truth, the substantial difference in vector field magnitude results in high VPE values of 641.5918 and 641.2468.} 
\label{fig:ldm_curlfree} 
\end{figure}

\autoref{tab:physical_evaluation_curlfree} shows the evaluation results for the irrotational vector fields. The MSE and RMSE results indicate that vector fields obtained by applying Helmholtz-Hodge decomposition (HHD) exhibit lower errors compared to those generated by LDM alone. Specifically, MSE improved from 0.063 to 0.036, and RMSE from 0.237 to 0.176. 
Moreover, SSIM improved significantly from 0.652 to 0.763.Furthermore, CME is a metric that indicates the magnitude of rotational components in vector fields, where lower values indicate irrotational characteristics. The evaluation results show that the CME value is $1.83\times10^{-19}$, which is remarkably close to zero compared to the initial vector field.
\autoref{fig:ldm_curlfree} shows the visualization results. The visualization reveals that the initial vector field generated by LDM exhibits a significant presence of unexpected harmonic components, as seen in \autoref{fig:ldm_curlfree}(a).
These results demonstrate that our proposed method enhances the vector field generation quality in terms of physical properties, numerical accuracy, and visual fidelity. 
The method improves performance over the LDM alone by removing irrelevant harmonic components from the initial vector field while maintaining irrotational properties.
Conversely, the VPE values exhibited minimal differences between LDM-only (90.56220) and HHD-applied (90.53893) cases due to the generated vector fields with relatively low vorticity.
Additionally, VPE values recorded high numbers for both LDM vector fields and the proposed method due to the lack of magnitude expressions in the sketches. 
As illustrated in \autoref{fig:ldm_curlfree}(b), the significant differences in vector field magnitudes resulted in high VPE values, even though the flow directions match.

\begin{figure}[t]
\centering 
\includegraphics[width=1\linewidth]{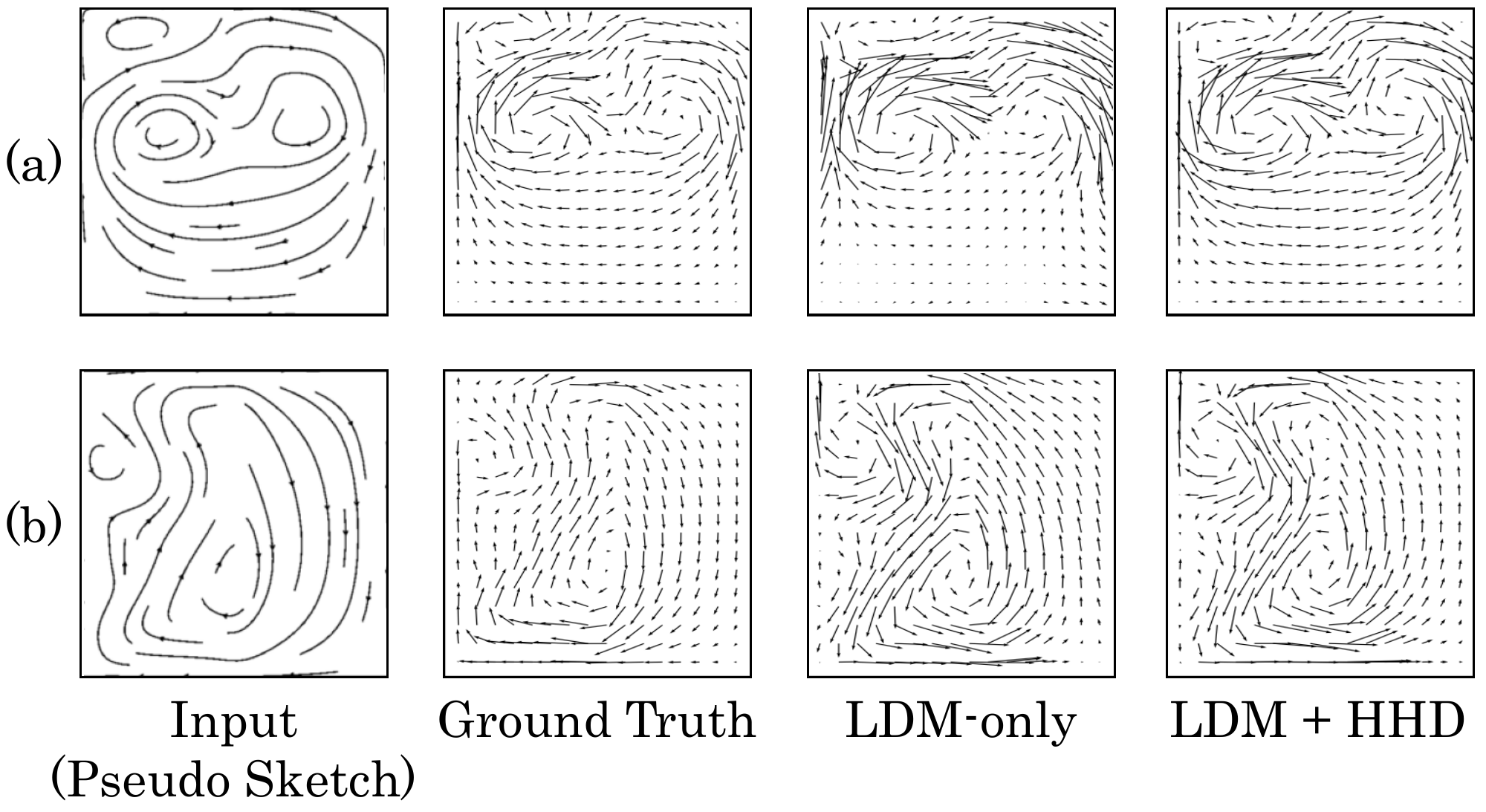} 
\caption{Examples of incompressible vector field generation and decomposition results: (a) An example where unnecessary components are removed. By extracting the divergence-free part using Helmholtz-Hodge decomposition (HHD), SSIM improves from 0.6493 to 0.8840, demonstrating a more accurate representation of the ground truth. (b) An example with significantly high SFE values. Although the generated vector field maintains a similar flow pattern, the presence of opposite vortices results in extremely high SFE values of 4077.2361 and 4077.1762.} \label{fig:ldm_divfree} 
\end{figure}

\autoref{tab:physical_evaluation_divfree} shows the evaluation results for incompressible vector fields. Vector fields processed with Helmholtz-Hodge decomposition show improved numerical accuracy compared to those generated by LDM alone. 
Specifically, MSE improved from 0.794 to 0.753, and RMSE from 0.770 to 0.735. SSIM also improved slightly from 0.432 to 0.451.
The CS value indicates the MSE of the vector field divergence component. It shows a value of $3.81\times10^{-35}$, which is substantially close to zero compared to the initial vector field.
\autoref{fig:ldm_divfree} shows the visualization results. 
As shown in \autoref{fig:ldm_divfree}(a), the vector field after applying Helmholtz-Hodge decomposition successfully generates a flow pattern similar to the ground truth by removing unnecessary components.
According to these results, our proposed method facilitates the generation of high-quality vector fields in terms of physical properties, numerical accuracy, and visual appearance.
However, SFE values remained unchanged between the initial vector fields generated by LDM and vector fields processed through Helmholtz-Hodge decomposition. This issue arises when the CS values of initial vector fields approach zero.
Furthermore, the high SFE values can be attributed to instances when LDM reconstructed vortices in opposite directions. As shown in \autoref{fig:ldm_divfree}(b), when vortices are reconstructed in opposite directions, SFE values increase significantly.
\subsection{Application Examples}
\begin{figure}
    \centering
    \includegraphics[width=1\linewidth]{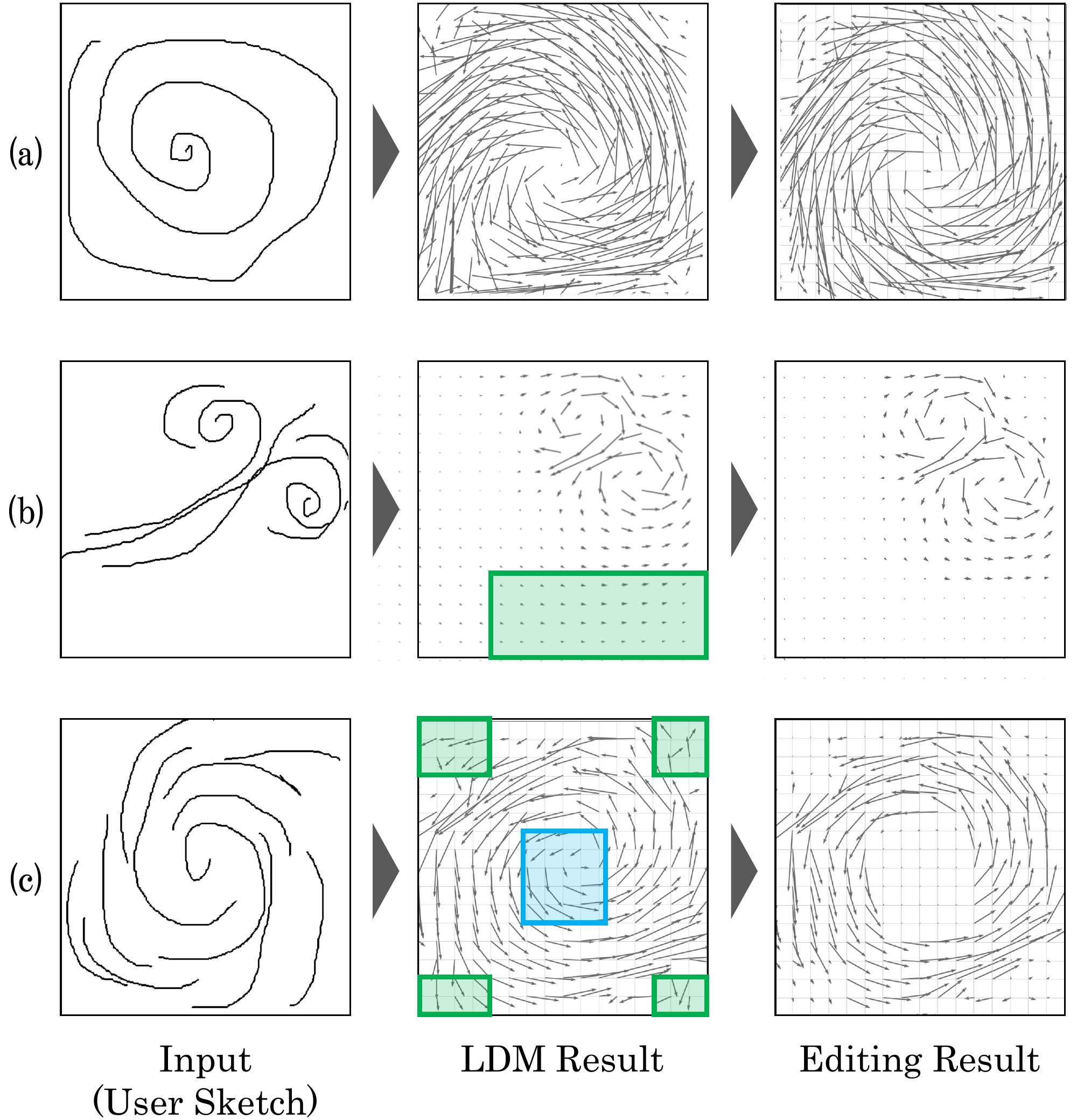}
    \caption{Examples of editing results for 2D vector fields generated from sketches. In examples (a) and (c), the input sketch corresponds to the swirl pattern, while in example (b), the input sketch corresponds to the plume pattern. The generated vector fields align with the shapes of the given sketches. In examples (b) and (c), we test with the editing mode. The editing results successfully match the activated operations. }
    \label{fig:enter-label}
\end{figure}
\autoref{fig:enter-label}(a) shows an example of designing a physically plausible vortex by extracting only the incompressible component from the LDM-based results.

\autoref{fig:enter-label}(b) shows an editing result of smoke motion. We removed the curl-free component from the entire LDM-based results and extracted only the incompressible component from the lower-right portion (green), resulting in a vector field that better reflects the sketch's intended behavior.

\autoref{fig:enter-label}(c) shows an example of complex editing result from sketch vortex flows. 
We first removed the harmonic component from the entire LDM-based vector field. 
Next, we extracted only the curl-free component from the central region (blue) to remove local rotation, and extracted only the incompressible component from the corner regions (green) to eliminate unwanted flows.

\section{Limitations and Future Work}

In our current vector field generation model based on a paired dataset, actual sketches may produce unintended local patterns or misaligned vector fields—for example, reversed flow directions or incorrect magnitudes. To address these issues, we plan to build a vector field generation model that is robust to sketches, by augmenting the dataset with real sketch data, following the approach of \cite{hu_sketch2vf_2019}, and revisiting our current pseudo-sketch generation pipeline. Furthermore, we plan to develop a method for explicitly embedding directional and magnitude information into the sketch data, enabling more stable and controllable vector field generation.

The current interface limits user-specified regions to rectangular shapes, making it difficult to extract physical properties from arbitrarily shaped areas such as circles, ellipses, or free-form curves. To overcome this limitation, we plan to incorporate the mesh-based Helmholtz-Hodge decomposition method proposed in \cite{farin_identifying_2003}, allowing for the extraction of physical properties from arbitrarily shaped regions. In addition, to handle the potential generation of unnatural vector fields along the boundaries of user-specified regions after decomposition, we are considering adopting a boundary velocity interpolation method based on energy minimization, following the approach proposed in \cite{sato_editing_2018}.
\begin{figure}
    \centering
    \includegraphics[width=1.0\linewidth]{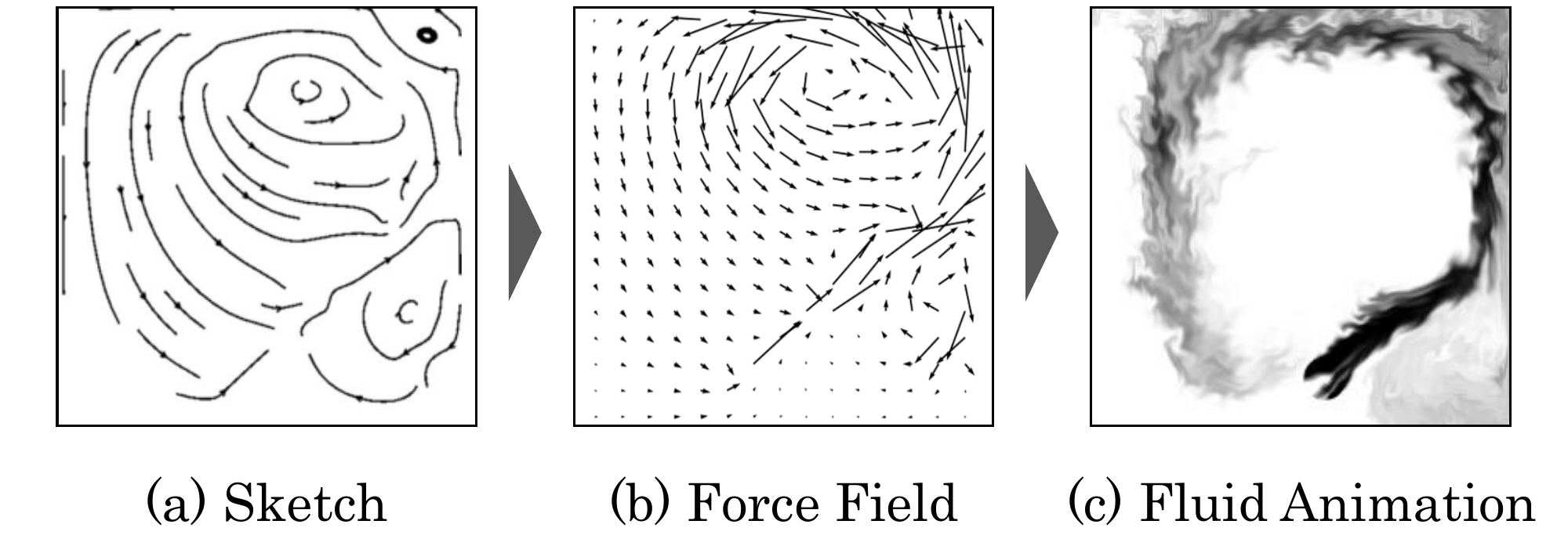}
    \caption{Application of generated vector fields to fluid simulation. The figure illustrates a smoke simulation result using vector fields produced by our proposed method.}
    \label{fig:fluid_simulation_application}
\end{figure}

Moreover, we will explore the potential of applying the generated vector fields as force fields in fluid simulations. ~\autoref{fig:fluid_simulation_application} shows an example of a smoke simulation result using vector fields generated by our method. In future work, we aim to integrate our approach with numerical fluid solvers such as the Navier-Stokes equations to enable intuitive sketch-based fluid control. We also plan to extend our method to 3D vector fields. Finally, to objectively evaluate the generation quality, we plan to conduct comparative experiments with other generative models \cite{hu_sketch2vf_2019, chang_sketch-guided_2024}.

\section{Conclusion}
We proposed a two-stage method for designing 2D vector fields while considering physical properties. We have combined an LDM-based sketch-guided vector field generation network with a physical property extraction interface utilizing Helmholtz-Hodge decomposition. Our system effectively handles physical properties that are not accounted for in conventional generative models. Additionally, several examples were presented to demonstrate the robustness of the proposed method. This framework holds significant potential for adaptation to address a broader range of fluid control problems in future research.

\section*{Acknowledgment}
This research result was based on the outcomes of the ``GENIAC (Generative AI Accelerator Challenge)'' project, implemented by the Ministry of Economy, Trade and Industry and the New Energy and Industrial Technology Development Organization (NEDO), aimed at strengthening domestic generative AI development capabilities. This work was supported by JSPS KAKENHI Grant Number 23K18514.

\bibliographystyle{st}
\bibliography{references}

\begin{thebibliography}{10}

\bibitem{wang_physics-based_2024}
Xiaokun Wang, Yanrui Xu, Sinuo Liu, Bo~Ren, Jiří Kosinka, Alexandru~C. Telea, Jiamin Wang, Chongming Song, Jian Chang, Chenfeng Li, Jian~Jun Zhang, and Xiaojuan Ban.
\newblock Physics-based fluid simulation in computer graphics: Survey, research trends, and challenges.
\newblock {\em Computational Visual Media~(CVMJ)}, Vol.~10, No.~5, pp. 803--858, 2024.

\bibitem{hu_sketch2vf_2019}
Zhongyuan Hu, Haoran Xie, Tsukasa Fukusato, Takahiro Sato, and Takeo Igarashi.
\newblock Sketch2VF: Sketch-based flow design with conditional generative adversarial network.
\newblock {\em Computer Animation and Virtual Worlds~(CAVW)}, Vol.~30, No.~3, pp. e1889:1--e1889:11, 2019.

\bibitem{xie2024dualsmoke}
Haoran Xie, Keisuke Arihara, Syuhei Sato, and Kazunori Miyata.
\newblock Dual{S}moke: Sketch-based smoke illustration design with two-stage generative model.
\newblock {\em Computational Visual Media~(CVMJ)}, pp. 1--15, 2024.

\bibitem{chang_diffsmoke_2025}
Hengyuan Chang, Tianyu Zhang, Syuhei Sato, and Haoran Xie.
\newblock {DiffSmoke}: {Two}-{Stage} {Sketch}-{Based} {Smoke} {Illustration} {Design} {Using} {Diffusion} {Models}.
\newblock {\em IEEE Access}, Vol.~13, pp. 44997--45009, 2025.

\bibitem{gregson_capture_2014}
James Gregson, Ivo Ihrke, Nils Thuerey, and Wolfgang Heidrich.
\newblock From capture to simulation: connecting forward and inverse problems in fluids.
\newblock {\em {ACM} Transactions on Graphics~(ToG)}, Vol.~33, No.~4, pp. 139:1--139:11, 2014.

\bibitem{pan_efficient_2017}
Zherong Pan and Dinesh Manocha.
\newblock Efficient solver for spacetime control of smoke.
\newblock {\em {ACM} Transactions on Graphics~(ToG)}, Vol.~36, No.~5, pp. 162:1--162:13, 2017.

\bibitem{xing_energy-brushes_2016}
Jun Xing, Rubaiat~Habib Kazi, Tovi Grossman, Li-Yi Wei, Jos Stam, and George Fitzmaurice.
\newblock Energy-brushes: Interactive tools for illustrating stylized elemental dynamics.
\newblock In {\em Proceedings of the 29th Annual Symposium on User Interface Software and Technology~(UIST)}, pp. 755--766. {ACM}, 2016.

\bibitem{ronneberger_u-net_2015}
Olaf Ronneberger, Philipp Fischer, and Thomas Brox.
\newblock U-{Net}: Convolutional networks for biomedical image segmentation.
\newblock {\em Lecture Notes in Computer Science~(LNCS)}, Vol. 9351, pp. 234--241, 2015.

\bibitem{rombach_high-resolution_2022}
Robin Rombach, Andreas Blattmann, Dominik Lorenz, Patrick Esser, and Bjorn Ommer.
\newblock High-resolution image synthesis with latent diffusion models.
\newblock In {\em Proceedings of {IEEE}/{CVF} Conference on Computer Vision and Pattern Recognition~({CVPR})}, pp. 10674--10685. {IEEE}, 2022.

\bibitem{hu_squeeze-and-excitation_2018}
Jie Hu, Li~Shen, Samuel Albanie, Gang Sun, and Enhua Wu.
\newblock Squeeze-and-excitation networks.
\newblock In {\em Proceedings of {IEEE}/{CVF} Conference on Computer Vision and Pattern Recognition~({CVPR})}, pp. 7132--7141. IEEE, 2018.

\bibitem{wang_image_2004}
Zhou Wang, A.C. Bovik, H.R. Sheikh, and E.P. Simoncelli.
\newblock Image quality assessment: from error visibility to structural similarity.
\newblock {\em IEEE Transactions on Image Processing~(TIP)}, Vol.~13, No.~4, pp. 600--612, 2004.

\bibitem{cao_teaching_2024}
Qinglong Cao, Ding Wang, Xirui Li, Yuntian Chen, Chao Ma, and Xiaokang Yang.
\newblock Teaching video diffusion model with latent physical phenomenon knowledge, November 2024.
\newblock arXiv:2411.11343 [cs].

\bibitem{farneback_two-frame_2003}
Gunnar Farnebäck.
\newblock Two-frame motion estimation based on polynomial expansion.
\newblock {\em Lecture Notes in Computer Science~(LNCS)}, Vol. 2749, pp. 363--370, 2003.

\bibitem{farin_identifying_2003}
Konrad Polthier and Eike Preu{\ss}.
\newblock Identifying vector field singularities Using a discrete hodge decomposition.
\newblock In {\em Visualization and Mathematics~{III}}, pp. 113--134. Springer, 2003.

\bibitem{sato_editing_2018}
Syuhei Sato, Yoshinori Dobashi, and Tomoyuki Nishita.
\newblock Editing fluid animation using flow interpolation.
\newblock {\em {ACM} Transactions on Graphics~(ToG)}, Vol.~37, No.~5, pp. 173:1--173:12, 2018.

\bibitem{chang_sketch-guided_2024}
Hengyuan Chang, Yichen Peng, Syuhei Sato, and Haoran Xie.
\newblock Sketch-guided flow field generation with diffusion model.
\newblock In {\em International Workshop on Advanced Imaging Technology ({IWAIT})}, Vol. 13164, pp. 290--295. {SPIE}, 2024.

\end{thebibliography}
\end{document}